\documentclass[prd,preprint,nofootinbib,tightenlines,showpacs,amsmath,amssymb,floatfix]{revtex4}

\def\be{\begin{equation}}
\def\ee{\end{equation}}
\def\bea{\begin{eqnarray}}
\def\eea{\end{eqnarray}}

\begin{document}

\title{Averaged null energy condition violation in a conformally flat
  spacetime}

\author{Douglas Urban}
\email{Douglas.Urban@tufts.edu}
\author{Ken D. Olum}
\email{kdo@cosmos.phy.tufts.edu}
\affiliation{Institute of
  Cosmology, Department of Physics and Astronomy, Tufts University,
  Medford, MA 02155, USA.}


\begin{abstract}
We show that the averaged null energy condition can be violated by a
conformally coupled scalar field in a conformally flat spacetime in
3+1 dimensions.  The violation is dependent on the quantum state and
can be made as large as desired.  It does not arise from the presence
of anomalies, although anomalous violations are also possible.  Since
all geodesics in conformally flat spacetimes are achronal, the
achronal averaged null energy condition is likewise violated.

\end{abstract}

\maketitle

\section{Introduction}

Without any restriction on the states of matter that can act as
sources, general relativity allows arbitrary spacetimes, which may
contain closed timelike curves, wormholes, and other such exotic
phenomena.  To prevent their occurrence requires restrictions on the
stress-energy tensor of matter, which are called \emph{energy
  conditions}.  For example, the usual classical fields obey the weak
energy condition: the energy density seen by any (timelike) observer
can never be negative.\footnote{Non-minimally-coupled scalar fields
  are an exception \cite{Barcelo:1999hq,Barcelo:2000zf,Fewster:2006ti}.  
  In this case the classical field can easily violate all pointwise energy
  conditions.  However, classical violations of ANEC
  \cite{Barcelo:1999hq,Barcelo:2000zf} are possible only
  if the field takes on Planck-scale values, which lead the effective
  Newton's constant to first diverge and then assume negative values.
  This may mean that such states are not physically realizable.}  From
this condition, wormholes, superluminal travel, and construction of time
machines can be ruled out
\cite{Friedman:1993ty,Olum:1998mu,Tipler:1976bi,Hawking:1991nk}.

Unfortunately, quantum fields can violate any restriction on the value
of the stress-energy tensor $T_{ab}$ at a point, so the above argument
does not hold in semiclassical gravity.  For example, a superposition
of the vacuum and a two-photon state gives negative energy density at
certain locations.  To make progress in
this case, one can go to averaged energy conditions which restrict
only certain averages of $T_{ab}$.  In particular, the exotic
situations mentioned above could be ruled out by the averaged null
energy condition (ANEC), which states that the projection of $T_{ab}$
onto the tangent vector of a null geodesic cannot give a negative
integral,
\be
\int_\gamma T_{ab} l^a l^b \ge 0
\ee
where $l^a$ is the tangent vector to the geodesic $\gamma$.

In Minkowski space, ANEC always holds
\cite{Klinkhammer:1991ki,Wald:1991xn}.  It cannot be violated even if
one allows arbitrary boundaries (generalizing the parallel plates of
the Casimir effect), as long as these do not approach arbitrarily
close to the geodesic \cite{Fewster:2006uf}.  The result of
\cite{Fewster:2006uf} applies also to spacetimes that are flat near
the geodesic but have curvature in distant places, as long as that
curvature does not change the causal structure near the geodesic.
However, this result does not apply for null geodesics that are
chronal, that is to say some of whose points are in the chronological
future of others.

A simple example of ANEC violation for chronal geodesics is given by
the Casimir-like system produced by compactifying one spatial
dimension in Minkowski space.  In this case both the energy density
and the pressure in the compactified direction are negative
everywhere, and ANEC is violated by geodesics going in the compact
direction.  Because of the compactification, all geodesics are
chronal.

ANEC can also be violated in 3+1 dimensional curved space.  An example
is given by the Schwarzschild spacetime in the Boulware vacuum state
\cite{Visser:1996iv}.  All complete geodesics (i.e., those that avoid the
singularity) violate ANEC, but all those geodesics are chronal.

In 1+1 dimensions, on the other hand, either all geodesics are chronal
(if the spatial dimension is compactified) or all geodesics are
achronal.  In the latter case ANEC always holds \cite{Wald:1991xn}.

The above considerations might lead one to guess that quantum fields
always obey ``achronal ANEC,'' i.e., they obey ANEC on any achronal
geodesic.  This condition is sufficient to rule out many exotic
situations \cite{Graham:2007va}, but even it is violated.  Visser
\cite{Visser:1994jb} showed that in general spacetimes, one can always
violate ANEC by rescaling (which does not change the chronality of a
geodesic) However, his violation results from the anomalous
transformation of the stress tensor.  It involves the logarithm of the
rescaling factor multiplied by a tiny number such as $1/(2880\pi^2)$.
Thus any realistic rescaling will have negligible effect.
Reference \cite{Graham:2007va} conjectured that a principle of
self-consistency could rule out this violation.  If one requires the
spacetime to be generated self-consistently by the state of the
quantum fields (perhaps with the addition of some classical matter),
Visser's anomalous violation would be too small to lead to the
curvature necessary to produce the violation.

In the next section, we exhibit a new curved-space ANEC
violation.  We study the conformally coupled scalar field in
conformally flat spacetimes and find a state-dependent violation which
can be made arbitrarily large.  The basic idea is to start with a
Minkowski-space quantum state which obeys ANEC (as it must), but
which violates the null energy condition (NEC), which requires that $T_{ab}
l^a l^b \ge 0$ for every point and every null vector $l^a$.  We then choose a
conformal transformation which enhances the NEC-violating regions in
the ANEC integral, so that in the conformally related spacetime ANEC
is violated.  By choosing appropriate quantum states, ANEC can be
violated to any desired degree.

The violation we discuss here differs from that of Visser
\cite{Visser:1994jb} in that it depends on the state, rather than
arising from anomalous terms, which depend only on the
spacetime curvature.  Additionally, the conformally flat systems
discussed here are a special case to which Visser's argument does not
apply, and where the scaling anomaly he discusses does not occur.

In Sec. III we construct another kind of ANEC violation arising
only from the anomalous terms in the transformation of the
stress-energy tensor, starting from the Minkowski-space vacuum.  The
violation depends on having an inhomogeneous conformal transformation,
rather than simply a rescaling, and does not depend the choice
of renormalization scale

In Sec. IV we review explicitly why these approaches cannot be used
to violate ANEC in 1+1 dimensions.  In this case the anomalous
contribution is always positive and cancels the largest effect that
can be generated by reweighting an NEC-violating Minkowski-space
state.

Finally, we conclude in Sec. V with some possibilities for how one
might rule out exotic phenomena even though ANEC does not always hold.

We work in units where $c=1$ and $\hbar = 1$.  Our sign conventions
are (+++) in the categorization of Misner, Thorne, and Wheeler
\cite{mtw}.

\section{Nonanomalous violation}\label{sec:violation}

We will construct our violation of ANEC as a conformal transformation
of a spacetime that obeys ANEC but violates NEC, i.e., there is a
geodesic $\gamma$ with tangent vector $l^a$, such that $T_{ab} l^a
l^b<0$ in certain places but $\int_\gamma T_{ab} l^a l^b \ge 0$.  For simplicity, we will take the untransformed spacetime to be Minkowski
space.  We will show that a conformal transformation can enhance the
contribution to the integral in those places where NEC is violated, so
that the overall integral is negative in the transformed spacetime.

We let our transformed metric be $\bar{g}_{ab}=\Omega^2(x)
\eta_{ab}$.  The stress-energy tensor then transforms as \cite{Birrellbook}
\be\label{eqn:Ttransformsimple}
\bar{T}_{ab}=\Omega^{-2}T_{ab}+ \text{anomaly}
\ee
The anomalous contribution depends only on local curvature terms and
is finite.  A null geodesic remains a null geodesic under a conformal
transformation, but the parameterization is no longer affine.  A new
affine parameterization is given by $d\bar\lambda=\Omega^2d\lambda$,
and so $\bar{l}^a = (dx^a/d\bar\lambda) = \Omega^{-2} l^a$.
The ANEC integral then becomes
\be\label{eqn:ANECtransform}
\int \bar{T}_{ab}\bar{l}^a \bar{l}^b d\bar{\lambda}=\int \Omega^{-4}T_{ab}l^a l^b d\lambda + \text{anomaly}\,.
\ee

For a given conformal transformation, we will exhibit a sequence of
states in which the nonanomalous term becomes arbitrarily negative.
Thus, even if the anomalous term is positive, there are states which
overcome it and make the ANEC integral negative.  In fact it is
possible to arrange the transformation so that the anomalous term also
gives a negative contribution.

Our argument follows closely the work of Fewster and Roman
\cite{FewsterRomanwitherratum} on null quantum inequalities.  A \emph{quantum
  inequality} is a restriction on the amount by which a weighted
average of the stress-energy tensor can be negative.  For example, for
a minimally coupled massless scalar in Minkowski space, we have
\cite{Ford:1994bj}
\be\label{eqn:timelikeQI}
\int \frac{\tau_0}{\pi(\tau^2+\tau_0^2)} T_{ab} V^a V^b d\tau\ge -
\frac{1}{32\pi^2\tau_0^4}
\ee
where the integral is taken over a timelike geodesic with tangent
vector $V^a$, parameterized by proper time
$\tau$, and $\tau_0$ is a arbitrary constant.  

Fewster and Roman \cite{FewsterRomanwitherratum} showed that no inequality such
as Eq.\ (\ref{eqn:timelikeQI}) can hold for null geodesics.
Specifically, for any affinely parameterized null geodesic
$\gamma(\lambda)$ with tangent vector $l^a$ and any smooth, bounded,
compactly-supported function $f(\lambda)$, Fewster and Roman construct
a sequence of states which make $\int f(\lambda) T_{ab}l^a l^b
d\lambda$ unboundedly negative.  We will use their construction to
produce a Minkowski-space state which will violate ANEC when
conformally transformed.

Consider a geodesic $\gamma$ as above and a smooth conformal transformation
$\Omega(x)$, with the properties that
\bea
&&\Omega(\gamma(\lambda))\le 1\,\text{everywhere on $\gamma$}\\
&&\text{$\Omega(\gamma(\lambda))$ is bounded from below by some $\epsilon > 0$}\\
&&\text{$\Omega(\gamma(\lambda))$ differs from 1 on a non-empty
  compact set of $\lambda$}
\eea
The conformal transformation shrinks the spacetime by some
bounded amount over some limited range of the geodesic.  We can then
define $g(\lambda) = \Omega(\gamma(\lambda))^{-4}$ and $f(\lambda) = g(\lambda) -1$,
and $f$ will then be smooth, bounded, and of compact support.

The ANEC integral in the conformally flat spacetime is
\be\label{eqn:conformalANEC}
\int \bar{T}_{ab}\bar{l}^a \bar{l}^b d\bar{\lambda}
= {\cal E} [g] + \text{anomaly}\,.
\ee
where ${\cal E}[g]$ is defined as the flat-spacetime integral with
sampling function $g$,
\be
{\cal E} [g] = \int_\gamma g(\lambda) T_{ab}l^a l^b d\lambda
\ee

Following \cite{FewsterRomanwitherratum}, we will now exhibit a sequence of
states $\psi_\alpha$ that will make the ANEC integral arbitrarily
negative.  Since we are concerned only with a counterexample to ANEC,
will not attempt to be general but opt instead for simplicity.  Our
procedure differs from that of \cite{FewsterRomanwitherratum} in that our
field is conformally rather than minimally coupled, and our
sampling function $g$ is not compactly supported but rather goes to 1
at large distances.

A massless field $\phi$ is defined by
\be
\phi(x)=\int\frac{d^3k}{(2\pi)^3(2\omega)^{1/2}}\left(a(k)e^{-ik_ax^a}+a^\dagger(k)e^{ik_ax^a}\right)\,.
\ee
We define a class of vacuum plus two particle state vectors, which
depend on a parameter $\alpha \in (0,1)$.  First, given the function
$f$, we will define a momentum parameter $\Lambda_0$ by
a procedure to be described later.  Then we define our states
\be\label{eqn:psi}
\psi_\alpha=N_\alpha \left[ |0> + \frac{\alpha^{-1/4}}{\Lambda^4}
\int_\Sigma \frac{d^3kd^3k'}{(2\pi)^3(2\pi)^3} \sqrt{kk'} |k,k'>  \right]
\ee
where $\Lambda = \Lambda_0/ \alpha$ is a momentum cutoff, $N_\alpha$
is a normalization constant,
\be
N_\alpha=\left( 1+\frac{\alpha^{3/2}}{128\pi^4} \right)^{-1/2}
\ee
and
\be
\int_\Sigma d^3k \quad\text{denotes}\quad \int_0^\Lambda k^2dk \int_{1-\alpha}^1 d\cos\theta
\int_0^{2\pi} d\phi
\ee
where $k$ is the magnitude of the vector ${\bf k}$, $\theta$ is the
angle between ${\bf k}$ and the tangent vector ${\bf l}$, and $\phi$
is the azimuthal angle.  These states excite only particles with momentum
less than $\Lambda$, and directed inside an angle
$\cos^{-1}(1-\alpha)$ from the null ray, which puts the
four-momentum inside a tightening and lengthening cone as $\alpha \to
0$.  Note that as $\alpha$ falls to zero, $N_\alpha \to 1$ and the
excitation term in Eq.~\eqref{eqn:psi} goes to zero. Thus the state
approaches the vacuum, but we shall see that its stress-energy tensor
does not.

In order to find the stress tensor, we need the normal ordered two
point function \cite{FewsterRomanwitherratum}
\be
\langle\psi_\alpha|:\phi(x)\phi(x'):|\psi_\alpha\rangle
= \frac{2N_\alpha^2}{\Lambda^4}
\int_\Sigma\frac{d^3kd^3k'}{(2\pi)^6}
\left[\alpha^{-1/4}e^{-i(k\cdot x+k'\cdot x')}
+\frac{\alpha^{1/2}}{8\pi^2}e^{i(-k\cdot x+k'\cdot x')}
\right]
\ee
The first term arises from the coupling of the two-particle states to
the vacuum.  The second arises from the coupling between the
two-particle states.  In the limit $\alpha\to0$, the first term is
dominant because the admixture of two-particle states becomes very
small.

The stress tensor for a conformally coupled scalar field is
\be
\label{eqn:stresstensor}
T_{ab}=\frac{2}{3}\phi_{;a}\phi_{;b} - \frac{1}{3}\phi_{;ab}\phi -
\frac{1}{6}g_{ab}g^{\rho\sigma}
\phi_{;\rho}\phi_{;\sigma} 
+ \frac{1}{12} g_{ab} \phi\Box\phi -
\frac{1}{6}\left[R_{ab} - \frac{1}{4}R
  g_{ab}\right]\phi^2
\ee
In flat space the curvature terms vanish, and terms
involving $g_{ab}$ vanish in the null projection, so
\be
l^al^bT_{ab}= \frac{2}{3}l^al^b\phi_{,a}\phi_{,b} - \frac{1}{3} l^al^b\phi_{,ab}\phi
\ee
We take the expectation value in the state $\psi_\alpha$ and
renormalize by subtracting the vacuum contribution (which is
equivalent to normal ordering), then set $x' = x$.   The first term becomes
\be
\frac{2}{3}<:\phi_{,a}\phi_{,b}l^a l^b:>_\alpha
= \frac{4N_\alpha^2}{3\Lambda^4}
\int_\Sigma\frac{d^3kd^3k'}{(2\pi)^6}
l^a k_a l^b k'_b
\left[-\alpha^{-1/4}e^{-ix\cdot(k+k')}
+\frac{\alpha^{1/2}}{8\pi^2}e^{ix\cdot(k-k')}
\right]
\ee
The other term is
\be
-\frac{1}{3}<:\phi_{,ab}\phi l^a l^b:>_\alpha=
\frac{2N_\alpha^2}{3\Lambda^4}
\int_\Sigma\frac{d^3kd^3k'}{(2\pi)^6}
(l^a k_a)^2
\left[\alpha^{-1/4}e^{-ix\cdot(k+k')}
+\frac{\alpha^{1/2}}{8\pi^2}e^{ix\cdot(k-k')}
\right]
\ee

We now specify a Fourier transform by
\be
\hat{f}(u) = \int dt e^{-iut}f(t)
\ee
Since $g(t) = f(t) +1$, $\hat{g}(u)=\hat{f}(u)+2\pi\delta(u)$.  From the
properties of $\Omega$, we see that $f$ is bounded and has a
well-defined, positive integral.  Thus $\hat f$ is continuous and
$\hat f(0)>0$.

For any fixed 4-vector $K$,
\be
\int d\lambda g(\lambda)e^{-i\gamma(\lambda)^a K_a}=\hat{g}(l \cdot K)
\ee
so we can write ${\cal E}[g]={\cal E}_1[g] + {\cal E}_2 [g]$, where
\be\label{eqn:g1}
{\cal E}_1[g] = \frac{N_\alpha^2\alpha^{1/2}}{12\pi^2\Lambda^4}
\int_\Sigma \frac{d^3kd^3k'}{(2\pi)^6} \left[(l \cdot k)^2 + 2 (l \cdot k)(l \cdot k')
\right]\hat{g}(l\cdot(k-k'))
\ee
\be\label{eqn:g2}
{\cal E}_2[g] = \frac{2N_\alpha^2\alpha^{-1/4}}{3\Lambda^4}
\int_\Sigma \frac{d^3kd^3k'}{(2\pi)^6} \left[(l \cdot k)^2 - 2 (l \cdot k)(l \cdot k')
\right]\hat{g}(l\cdot(k+k'))
\ee

We will first consider ${\cal E}_2[f]$, following
\cite{FewsterRomanwitherratum}.  Since we are in flat space, the tangent
vector $l$ is constant.  We can take
it to have unit time component, so that $k\cdot l = k (1-\cos\theta)$.
We do the azimuthal integrations and change
variables to $v = k\alpha$, $u = k\cdot l$, and
similarly for $v'$ and $u'$.  We find
\be\label{eqn:f2}
{\cal E}_2[f] = \frac{2N_\alpha^2\alpha^{-1/4}}{3(2\pi)^4\Lambda_0^4}
\int_0^{\Lambda_0} dv \int_0^{\Lambda_0} dv' v v'
\int_0^v du \int_0^{v'} du' 
[u^2 - 2 u u'] \hat{f}(u+u')
\ee
Now $\hat f > 0$.  Since $\hat f$ is continuous, we can choose
$\Lambda_0>0$ such that $\hat f(u)$ is arbitrarily close to $\hat
f(0)$.  Thus we can make the integrals in Eq.~(\ref{eqn:f2})
arbitrarily close to
\be
\hat f(0)\int_0^{\Lambda_0} dv \int_0^{\Lambda_0} dv' v v'
\int_0^v du \int_0^{v'} du' 
[u^2 - 2 u u']
= -\frac{13}{1440}\hat f(0) < 0
\ee
As $\alpha\to 0$, the prefactor in Eq.~(\ref{eqn:f2}) goes to positive
infinity, so we conclude that ${\cal E}_2[f]\to-\infty$ in this limit.

The rest of the terms are all finite.  Equation (\ref{eqn:g1}) gives
\be\label{eqn:f1}
{\cal E}_1[f]=\frac{N_\alpha^2\alpha^{1/2}}{12\pi^2\Lambda_0^4}
\int_0^{\Lambda_0} dv \int_0^{\Lambda_0} dv' v v'
\int_0^v du \int_0^{v'} du' 
[u^2 + 2 u u'] \hat{f}(u-u')
\ee
Since $f$ has compact support, $\hat f$ is bounded and the
integrals give some finite number independent of $\alpha$.  Since the
power of $\alpha$ is positive in this case, we find that ${\cal
  E}_1[f]\to0$ as $\alpha\to 0$.

In addition we have the delta function in Eqs.\ (\ref{eqn:g1},\ref{eqn:g2}),
which gives the flat-spacetime ANEC integral discussed in Sec. II D of
Ref.\ \cite{FewsterRomanwitherratum}.  Since $k$ is restricted to a cone around
the direction of $l$, $l\cdot k\ge 0$.  There is no contribution
to ${\cal E}_2[\delta]$ except from $k=k'=0$, in which case the term in
brackets vanishes.  Thus ${\cal E}_2[\delta]= 0$.

Finally we have
\be
{\cal E}_1[\delta]=\frac{N_\alpha^2\alpha^{1/2}}{12\pi^2\Lambda_0^4}
\int_0^{\Lambda_0} dv \int_0^{\Lambda_0} dv' v v'
\int_0^v du \int_0^{v'} du' 
[u^2 + 2 u u'] \delta(u-u')
\ee
Again the integrals give a finite number, and the prefactor goes to
zero, so ${\cal E}_1[f]\to0$ as $\alpha\to 0$, and finally
\be
\lim_{\alpha\to0} {\cal E}[g]\to -\infty
\ee

Thus for a spacetime given by fixed conformal transformation
$\Omega$, we can find a quantum state such that ${\cal E}[g]$ is
arbitrarily negative.  In particular, any positive anomalous term can
be overcome by large negative ${\cal E}[g]$, so that
Eq.~(\ref{eqn:conformalANEC}) is negative and ANEC is violated.

\section{Anomalous Violation}\label{sec:anomalous}
The prior example constructs a violation of ANEC over a class of
excited states. The contribution from the transformed $T_{\mu\nu}$
dominates the anomalous terms. It is also possible to construct a
spacetime where the anomalous term is negative, and thus even for the
vacuum state, with $T_{\mu\nu}=0$, a violation can occur. In addition
to the example found by Visser, we find cases which are conformally
flat. In these, there is no dependence at all on the renormalization
scale $\mu$, only the geometry of the new space. 

Conformal transformation properties are taken from
\cite{Birrellbook,WaldGRbook}. The transformation is
$\bar{g}_{ab}=\Omega^2 g_{ab}$. In the following, derivatives of
barred quantities are always meant to be taken in the new metric, and
unbarred quantities in the original metric. Derivatives of the
transformation function $\Omega$ are also taken in the old
coordinates.  

The stress tensor is given by Eq.~\eqref{eqn:stresstensor}. For a
conformally coupled scalar field the transformation properties are
known \cite{Page:1982fm}. We specialize to the case where the initial
spacetime is Minkowski. Thus, curvature quantities in the
untransformed spacetime all vanish, and the Weyl tensor vanishes even
in the transformed spacetime. We have the particular
transformation
 \be 
 \label{eqn:Ttransform}
 \bar{T}^a_b= \Omega^{-4}T^a_b
  - 2 \beta \bar{H}^a_b - \frac{\gamma}{6} \bar{I}^a_b 
\ee
The constants are dependent on the spin of the field; for a real scalar
field,
\begin{eqnarray}
\beta &=& -\frac{1}{5760 \pi^2} \\
\gamma &=& -2\beta
\end{eqnarray} 
The tensors $\bar H$ and $\bar I$ are given by
\begin{eqnarray}
\bar{H}_{ab} = -\bar{R}^c_a
\bar{R}_{c b} + \frac{2}{3}\bar{R}\bar{R}_{ab} +
\left(\frac{1}{2}\bar{R}^c_d \bar{R}^d_c - \frac{1}{4}\bar{R}^2
\right) \bar{g}_{ab} \\
\label{I}
\bar{I}_{ab} = 2\bar{R}_{;ab} - 2\bar{R}\bar{R}_{ab} + \left(\frac{1}{2}\bar{R}^2 -
2\Box \bar{R}\right) \bar{g}_{ab} 
\end{eqnarray}

We eliminate those terms which will not contribute due to the null
projection as well as the $R_{;ab}$ in \eqref{I} which appears in ANEC as $l^b(R_{,a}l^a)_{,b}$ and thus vanishes upon integration. After this, and expressing $\gamma$ in terms of $\beta$, we find
\be
\bar{T}_{ab}=\Omega^{-2}T_{ab} +
2\beta\left[\bar{R}^c_a\bar{R}_{cb} -
  \bar{R}\bar{R}_{ab} \right]
\ee
The curvatures in the new spacetime are given by
\begin{eqnarray}
\bar{R}_{cb} &=& - 2\omega_{,cb} -
g_{cb}\Box\omega + 2\omega_{,b}\omega_{,c} -
2g_{cb}\omega_{,\rho}\omega^{,\rho} \\
\bar{R}&=&\Omega^{-2}\left[
  -6\Box\omega-6\omega_{,\rho}\omega^{,\rho}\right]
\end{eqnarray}
with $\omega=\ln\Omega$.  Again dropping terms with $g_{ab}$, the stress tensor is given by
\bea
\bar{T}_{ab}=\Omega^{-2}T_{ab} + 8\beta\Omega^{-2} \big[ {\omega^{,c}}_a\omega_{,cb} -2
  \left(\Box\omega+\omega^{,c}\omega_{,c}\right)\left(\omega_{,ab}-\omega_{,a}\omega_{,b}\right) \nonumber \\
   - \omega^{,c}\omega_{,a}\omega_{,cb} - \omega^{,c}\omega_{,b}\omega_{,ca} \big]
\eea

Now we give a specific example of a transformation which violates
ANEC. We take an initial state with $T_{\mu\nu}=0$, so the state does
not contribute to $\bar T_{\mu\nu}$. We will work in Minkowski space
in null coordinates, with $u=(z-t)/\sqrt{2}$ and
$v=(z+t)/\sqrt{2}$. We take our geodesic going in the $v$ direction
along the line $u=x=y=0$. We choose the
particular transformation
\be
\omega=(a+bx^2r^{-2})e^{-(u^2+v^2+x^2+y^2)/r^2}
\ee
This gives a localized transformation, so our spacetime is both
conformally and asymptotically flat. We take $a$ and $b$ both much
less than one, so we may ignore terms of order $\omega^3$. That leaves
us with only
\be
\bar{T}_{vv}=8\beta\Omega^{-2} \big[g^{cd}\omega_{,cv}\omega_{,dv} -2
 \Box\omega\omega_{,vv}  \big]
\ee
The first term vanishes because $\omega_{,cv}=0$ unless the index $c$
is $v$, but $g^{vv}=0$.  The remaining term is
the product of
\be
\Box\omega=2r^{-2}\left(b-2a\right)e^{-v^2/r^2}
\ee
and
\be
\omega_{,vv}= 2ar^{-2}\left( 2v^2r^{-2}-1\right)e^{-v^2/r^2}
\ee
Now together we have
\be
\bar{T}_{vv}= -64\beta ar^{-4} \left(
2v^2r^{-2}-1\right)\left(b-2a\right)e^{-2v^2/r^2}
\ee
Integrating over the full geodesic gives
\be
\int_{-\infty}^{+\infty}
\bar{T}_{vv}dv=\left(ab-2a^2\right)\frac{16\beta\sqrt{2\pi}}{r^3}
\ee
The constant $\beta < 0$. We can choose $1\gg b>2a$ so that the ANEC
integral is negative.

\section{1+1 Dimensions}

It is interesting to compare the results of previous sections with the
situation in 1+1 dimensions.  In that case we know \cite{Wald:1991xn}
that ANEC cannot be violated, even in curved space.  What happens when
we attempt to violate it using the techniques of previous sections?

First, our construction of a Minkowski-space state that violates a
weighted average of NEC depended on a cone of momenta surrounding the
tangent vector to our null geodesic.  In 1+1 dimensions, there are no
transverse directions, so that technique cannot work.  In fact, unlike
in 3+1 dimensions, there is a 1+1-dimensional quantum inequality
derived by Flanagan \cite{Flanagan:1997gn},
\be\label{eqn:Flanagan}
{\cal E} [g] = \int_\gamma T_{ab} l^a l^b d\lambda\geq - \frac{1}{48\pi} \int_\gamma \frac{g'(\lambda)^2}{g(\lambda)}d\lambda
\ee
for any smooth, non-negative function $g$.

Nevertheless, Eq.~(\ref{eqn:Flanagan}) still permits NEC violation,
and we can still enhance that violation.  In 1+1 dimensions,
Eq.~(\ref{eqn:Ttransformsimple}) becomes
\be\label{eqn:Ttransform2}
\bar{T}_{ab}=T_{ab}+ \text{anomaly}
\ee
and the ANEC integral, Eq.~(\ref{eqn:ANECtransform}), becomes
\be\label{eqn:ANECtransform2}
\int \bar{T}_{ab}\bar{l}^a \bar{l}^b d\bar{\lambda}=\int \left(T_{ab}+ \text{anomaly}\right)\Omega^{-2}l^a l^b d\lambda ,.
\ee
Thus if NEC is violated in certain locations in Minkowski space, we
can choose $\Omega \ll 1$ there to enhance their contribution to
Eq.~(\ref{eqn:ANECtransform2}).  However, we cannot make this
contribution arbitrarily large by the choice of states, because the NEC
violation is restricted by Eq.~(\ref{eqn:Flanagan}).

The anomalous term in Eq.~(\ref{eqn:Ttransform2}) (see
Eq.~(6.134) of \cite{Birrellbook}) is
\be
\frac{1}{12\pi}\left[\Omega^{-1}\Omega_{,ab}
 -2\Omega^{-2}\Omega_{,a}\Omega_{,b}
+g_{ab}g^{cd}\left((3/2)\Omega^{-2}\Omega_{,b}\Omega_{,c}\Omega^{-1}
-\Omega_{,bc}\right)\right]
\ee
The term proportional to $g_{ab}$ does not contribute to NEC so
\be
\int \bar{T}_{ab}\bar{l}^a \bar{l}^b d\bar{\lambda}=
\int \left[\Omega^{-2}T_{ab} + \frac{1}{12\pi}\left(\Omega^{-3}\Omega_{,ab}
 -2\Omega^{-4}\Omega_{,a}\Omega_{,b}\right)\right]l^a l^b d\lambda
\ee
We can integrate the anomalous terms by parts.  We write
\be
\left(\Omega^{-3}\Omega_{,a}\right)_{,b} = \Omega^{-3}\Omega_{,ab}
- 3 \Omega^{-4}\Omega_{,a}\Omega_{,b}
\ee
In our situation, $\Omega\to 1$ as $\lambda\to \pm\infty$, and thus
$\Omega_{,a}\to0$ in that limit.  So the total derivative does not
contribute, and
\be\label{eqn:finalTbar2}
\int \bar{T}_{ab}\bar{l}^a \bar{l}^b d\bar{\lambda}=
\int \left(\Omega^{-2}T_{ab}
 +\frac{1}{12\pi}\Omega^{-4}\Omega_{,a}\Omega_{,b}\right)l^a l^b d\lambda
\ee
The anomalous term in Eq.~(\ref{eqn:finalTbar2}) is manifestly
positive, so in 1+1 dimensions there is no anomalous violation as in
Sec. \ref{sec:anomalous}.   In fact, when we define $g(\lambda)
=\Omega(\gamma(\lambda))^{-2}$, we find the anomalous contribution is just
\be
\frac{1}{48\pi} \int\frac{g'(\lambda)^2}{g(\lambda)}d\lambda
\ee
and so by Eq.~(\ref{eqn:Flanagan}),
\be
\int \bar{T}_{ab}\bar{l}^a \bar{l}^b d\bar{\lambda} \ge 0\,;
\ee
ANEC is always obeyed.

This derivation is essentially the same used by Flanagan
\cite{Flanagan:2002bd} to generalize his quantum inequality,
Eq.~(\ref{eqn:Flanagan}), to curved spacetimes.  We find it remarkable
that Eq.~(\ref{eqn:Flanagan}), which is a statement entirely about
quantum field theory in flat spacetime somehow ``knows'' about
the anomalous transformation of $T_{ab}$ in such a way that
they together preserve ANEC in curved spacetime.

\section{Conclusion}

We have given two explicit violations of the achronal averaged null
energy condition, both in spaces which are conformally and
asymptotically flat. First we used a transformation which amplifies
the NEC-violating portions of a sequence of excited states. As the
momentum grows in magnitude and is constrained within a cone which
increasingly narrows around the direction of the null geodesic, the
ANEC integral becomes increasingly negative. This effect can be seen
in a broad class of states and transformations; we gave a specific
example for concreteness.

The second violation was constructed purely from the geometric
anomalous terms in the stress tensor. In this way we find negative
average energy in some conformally flat spaces even in the vacuum
state. As the deviation from flat space becomes more sharply
localized, the violation grows. Both of these violations can become
arbitrarily negative. 

We now wonder if there is any possibility to exclude exotic phenomena
from general relativity with some weaker condition that would not be
violated by quantum fields.  One option is requiring an additional
transverse average over a congruence of geodesics. Physically this is
a natural requirement, as any nonzero-sized exotic feature, such as a
wormhole or time machine, would require some certain level of ANEC
violation over some nonzero range of geodesics.

It appears that both of the above violations could be softened by some
transverse averaging. In the example of Sec. \ref{sec:violation},
the stress-energy tensor oscillates rapidly in the transverse
direction \cite{FewsterRomanwitherratum}.  Likewise, the violation in
Sec. \ref{sec:anomalous} grows as $r\to0$, where $r$ parameterizes
the width of the deviation from flatness. Averaging over a distance
greater than $r$ could cancel this effect. Timelike averages of null
quantities have been considered in \cite{FewsterRomanwitherratum}; one
could also consider spacetime averages.

Another possibility is the additional requirement of self-consistency,
that is, that the field and geometry be a solution to the
semiclassical Einstein equation $G_{\mu\nu}=8\pi\left<
T_{\mu\nu}\right>$, where $T_{\mu\nu}$ is the stress tensor of some
state of a set of fields in the background whose Einstein tensor is
$G_{\mu\nu}$.  In both the above examples we have computed the
stress tensor in a given background without attempting to impose self-consistency. Progress along this line has been made in
\cite{Flanagan:1996gw} for perturbations of flat space.
Ref.~\cite{Fewster:2007ec} finds state-dependent bounds on averaged
energies, which may also be useful in this context.  It is possible that
self-consistency may be enough to enforce the energy conditions in the
general case \cite{Graham:2007va}.

\section*{Acknowledgments}

The authors thank C. Fewster, L. Ford, N. Graham, and T.
Roman for helpful discussions.  This research was supported in part by
grants RFP1-06-024 and RFP2-06-23 from The Foundational Questions
Institute (fqxi.org).


\begin{thebibliography}{22}
\expandafter\ifx\csname natexlab\endcsname\relax\def\natexlab#1{#1}\fi
\expandafter\ifx\csname bibnamefont\endcsname\relax
  \def\bibnamefont#1{#1}\fi
\expandafter\ifx\csname bibfnamefont\endcsname\relax
  \def\bibfnamefont#1{#1}\fi
\expandafter\ifx\csname citenamefont\endcsname\relax
  \def\citenamefont#1{#1}\fi
\expandafter\ifx\csname url\endcsname\relax
  \def\url#1{\texttt{#1}}\fi
\expandafter\ifx\csname urlprefix\endcsname\relax\def\urlprefix{URL }\fi
\providecommand{\bibinfo}[2]{#2}
\providecommand{\eprint}[2][]{\url{#2}}

\bibitem[{\citenamefont{Barcelo and Visser}(1999)}]{Barcelo:1999hq}
\bibinfo{author}{\bibfnamefont{C.}~\bibnamefont{Barcelo}} \bibnamefont{and}
  \bibinfo{author}{\bibfnamefont{M.}~\bibnamefont{Visser}},
  \bibinfo{journal}{Phys. Lett.} \textbf{\bibinfo{volume}{B466}},
  \bibinfo{pages}{127} (\bibinfo{year}{1999}), \eprint{gr-qc/9908029}.

\bibitem[{\citenamefont{Barcelo and Visser}(2000)}]{Barcelo:2000zf}
\bibinfo{author}{\bibfnamefont{C.}~\bibnamefont{Barcelo}} \bibnamefont{and}
  \bibinfo{author}{\bibfnamefont{M.}~\bibnamefont{Visser}},
  \bibinfo{journal}{Class. Quant. Grav.} \textbf{\bibinfo{volume}{17}},
  \bibinfo{pages}{3843} (\bibinfo{year}{2000}), \eprint{gr-qc/0003025}.

\bibitem[{\citenamefont{Fewster and Osterbrink}(2006)}]{Fewster:2006ti}
\bibinfo{author}{\bibfnamefont{C.~J.} \bibnamefont{Fewster}} \bibnamefont{and}
  \bibinfo{author}{\bibfnamefont{L.~W.} \bibnamefont{Osterbrink}},
  \bibinfo{journal}{Phys. Rev.} \textbf{\bibinfo{volume}{D74}},
  \bibinfo{pages}{044021} (\bibinfo{year}{2006}), \eprint{gr-qc/0606009}.

\bibitem[{\citenamefont{Friedman et~al.}(1993)\citenamefont{Friedman, Schleich,
  and Witt}}]{Friedman:1993ty}
\bibinfo{author}{\bibfnamefont{J.~L.} \bibnamefont{Friedman}},
  \bibinfo{author}{\bibfnamefont{K.}~\bibnamefont{Schleich}}, \bibnamefont{and}
  \bibinfo{author}{\bibfnamefont{D.~M.} \bibnamefont{Witt}},
  \bibinfo{journal}{Phys. Rev. Lett.} \textbf{\bibinfo{volume}{71}},
  \bibinfo{pages}{1486} (\bibinfo{year}{1993}), \eprint{gr-qc/9305017}.

\bibitem[{\citenamefont{Olum}(1998)}]{Olum:1998mu}
\bibinfo{author}{\bibfnamefont{K.~D.} \bibnamefont{Olum}},
  \bibinfo{journal}{Phys. Rev. Lett.} \textbf{\bibinfo{volume}{81}},
  \bibinfo{pages}{3567} (\bibinfo{year}{1998}), \eprint{gr-qc/9805003}.

\bibitem[{\citenamefont{Tipler}(1976)}]{Tipler:1976bi}
\bibinfo{author}{\bibfnamefont{F.~J.} \bibnamefont{Tipler}},
  \bibinfo{journal}{Phys. Rev. Lett.} \textbf{\bibinfo{volume}{37}},
  \bibinfo{pages}{879} (\bibinfo{year}{1976}).

\bibitem[{\citenamefont{Hawking}(1992)}]{Hawking:1991nk}
\bibinfo{author}{\bibfnamefont{S.~W.} \bibnamefont{Hawking}},
  \bibinfo{journal}{Phys. Rev.} \textbf{\bibinfo{volume}{D46}},
  \bibinfo{pages}{603} (\bibinfo{year}{1992}).

\bibitem[{\citenamefont{Klinkhammer}(1991)}]{Klinkhammer:1991ki}
\bibinfo{author}{\bibfnamefont{G.}~\bibnamefont{Klinkhammer}},
  \bibinfo{journal}{Phys. Rev.} \textbf{\bibinfo{volume}{D43}},
  \bibinfo{pages}{2542} (\bibinfo{year}{1991}).

\bibitem[{\citenamefont{Wald and Yurtsever}(1991)}]{Wald:1991xn}
\bibinfo{author}{\bibfnamefont{R.~M.} \bibnamefont{Wald}} \bibnamefont{and}
  \bibinfo{author}{\bibfnamefont{U.}~\bibnamefont{Yurtsever}},
  \bibinfo{journal}{Phys. Rev.} \textbf{\bibinfo{volume}{D44}},
  \bibinfo{pages}{403} (\bibinfo{year}{1991}).

\bibitem[{\citenamefont{Fewster et~al.}(2007)\citenamefont{Fewster, Olum, and
  Pfenning}}]{Fewster:2006uf}
\bibinfo{author}{\bibfnamefont{C.~J.} \bibnamefont{Fewster}},
  \bibinfo{author}{\bibfnamefont{K.~D.} \bibnamefont{Olum}}, \bibnamefont{and}
  \bibinfo{author}{\bibfnamefont{M.~J.} \bibnamefont{Pfenning}},
  \bibinfo{journal}{Phys. Rev.} \textbf{\bibinfo{volume}{D75}},
  \bibinfo{pages}{025007} (\bibinfo{year}{2007}), \eprint{gr-qc/0609007}.

\bibitem[{\citenamefont{Visser}(1996)}]{Visser:1996iv}
\bibinfo{author}{\bibfnamefont{M.}~\bibnamefont{Visser}},
  \bibinfo{journal}{Phys. Rev.} \textbf{\bibinfo{volume}{D54}},
  \bibinfo{pages}{5116} (\bibinfo{year}{1996}), \eprint{gr-qc/9604008}.

\bibitem[{\citenamefont{Graham and Olum}(2007)}]{Graham:2007va}
\bibinfo{author}{\bibfnamefont{N.}~\bibnamefont{Graham}} \bibnamefont{and}
  \bibinfo{author}{\bibfnamefont{K.~D.} \bibnamefont{Olum}},
  \bibinfo{journal}{Phys. Rev.} \textbf{\bibinfo{volume}{D76}},
  \bibinfo{pages}{064001} (\bibinfo{year}{2007}), \eprint{0705.3193}.

\bibitem[{\citenamefont{Visser}(1995)}]{Visser:1994jb}
\bibinfo{author}{\bibfnamefont{M.}~\bibnamefont{Visser}},
  \bibinfo{journal}{Phys. Lett.} \textbf{\bibinfo{volume}{B349}},
  \bibinfo{pages}{443} (\bibinfo{year}{1995}), \eprint{gr-qc/9409043}.

\bibitem[{\citenamefont{Misner, Thorne and Wheeler}(1973)}]{mtw}
\bibinfo{author}{\bibfnamefont{C.~W.} \bibnamefont{Misner}}, 
  \bibinfo{author}{\bibfnamefont{K.~S.} \bibnamefont{Thorne}} \bibnamefont{and}
  \bibinfo{author}{\bibfnamefont{J.~A.} \bibnamefont{Wheeler}},
  \emph{\bibinfo{title}{{Gravitation}}}
  (\bibinfo{publisher}{W. H. Freeman and Company}, \bibinfo{year}{1973}).

\bibitem[{\citenamefont{Birrell and Davies}(1982)}]{Birrellbook}
\bibinfo{author}{\bibfnamefont{N.~D.} \bibnamefont{Birrell}} \bibnamefont{and}
  \bibinfo{author}{\bibfnamefont{P.~C.~W.} \bibnamefont{Davies}},
  \emph{\bibinfo{title}{{Quantum Fields in Curved Space}}}
  (\bibinfo{publisher}{Cambridge University Press}, \bibinfo{year}{1982}).

\bibitem[{\citenamefont{Fewster and Roman}(2003)}]{FewsterRomanwitherratum}
\bibinfo{author}{\bibfnamefont{C.~J.} \bibnamefont{Fewster}} \bibnamefont{and}
  \bibinfo{author}{\bibfnamefont{T.~A.} \bibnamefont{Roman}},
  \bibinfo{journal}{Phys. Rev.} \textbf{\bibinfo{volume}{D67}},
  \bibinfo{pages}{044003} (\bibinfo{year}{2003}), \bibinfo{note}{erratum: {\bf
  D80}, 069903(E)}.

\bibitem[{\citenamefont{Ford and Roman}(1995)}]{Ford:1994bj}
\bibinfo{author}{\bibfnamefont{L.~H.} \bibnamefont{Ford}} \bibnamefont{and}
  \bibinfo{author}{\bibfnamefont{T.~A.} \bibnamefont{Roman}},
  \bibinfo{journal}{Phys. Rev.} \textbf{\bibinfo{volume}{D51}},
  \bibinfo{pages}{4277} (\bibinfo{year}{1995}), \eprint{gr-qc/9410043}.

\bibitem[{\citenamefont{Wald}(1984)}]{WaldGRbook}
\bibinfo{author}{\bibfnamefont{R.~M.} \bibnamefont{Wald}},
  \emph{\bibinfo{title}{General Relativity}} (\bibinfo{publisher}{Chicago
  University Press}, \bibinfo{year}{1984}).

\bibitem[{\citenamefont{Page}(1982)}]{Page:1982fm}
\bibinfo{author}{\bibfnamefont{D.~N.} \bibnamefont{Page}},
  \bibinfo{journal}{Phys. Rev.} \textbf{\bibinfo{volume}{D25}},
  \bibinfo{pages}{1499} (\bibinfo{year}{1982}).

\bibitem[{\citenamefont{Flanagan}(1997)}]{Flanagan:1997gn}
\bibinfo{author}{\bibfnamefont{E.~E.} \bibnamefont{Flanagan}},
  \bibinfo{journal}{Phys. Rev.} \textbf{\bibinfo{volume}{D56}},
  \bibinfo{pages}{4922} (\bibinfo{year}{1997}), \eprint{gr-qc/9706006}.

\bibitem[{\citenamefont{Flanagan}(2002)}]{Flanagan:2002bd}
\bibinfo{author}{\bibfnamefont{E.~E.} \bibnamefont{Flanagan}},
  \bibinfo{journal}{Phys. Rev.} \textbf{\bibinfo{volume}{D66}},
  \bibinfo{pages}{104007} (\bibinfo{year}{2002}), \eprint{gr-qc/0208066}.

\bibitem[{\citenamefont{Flanagan and Wald}(1996)}]{Flanagan:1996gw}
\bibinfo{author}{\bibfnamefont{E.~E.} \bibnamefont{Flanagan}} \bibnamefont{and}
  \bibinfo{author}{\bibfnamefont{R.~M.} \bibnamefont{Wald}},
  \bibinfo{journal}{Phys. Rev.} \textbf{\bibinfo{volume}{D54}},
  \bibinfo{pages}{6233} (\bibinfo{year}{1996}), \eprint{gr-qc/9602052}.

\bibitem[{\citenamefont{Fewster and Osterbrink}(2008)}]{Fewster:2007ec}
\bibinfo{author}{\bibfnamefont{C.~J.} \bibnamefont{Fewster}} \bibnamefont{and}
  \bibinfo{author}{\bibfnamefont{L.~W.} \bibnamefont{Osterbrink}},
  \bibinfo{journal}{J. Phys.} \textbf{\bibinfo{volume}{A41}},
  \bibinfo{pages}{025402} (\bibinfo{year}{2008}), \eprint{0708.2450}.

\end{thebibliography}
\end{document}